\documentclass[12pt]{article}

\def\be{\begin{equation}}
\def\ee{\end{equation}}
\def\ba{\begin{eqnarray}}
\def\ea{\end{eqnarray}}

\def\ra{\rangle}

\def\h{\hskip 1cm}

\def\lo{\longrightarrow}

\usepackage{graphicx}
\begin{document}
\begin{titlepage}
\vspace{4cm}
\begin{center}{\Large \bf Reply to "Comment on quantum secret sharing based on
reusable Greenbergr-Horne-Zeilinger states
as secure carriers"}\\
\vspace{2cm}
 V. Karimipour \footnote{email:vahid@sharif.edu}
\vspace{1cm} \\ Department of Physics, Sharif University of Technology,\\
P.O. Box 11365-9161,\\ Tehran, Iran
\end{center}
\vskip 3cm
\begin{abstract}
In a recent comment, it has been shown that in a quantum secret
sharing protocol proposed in [S. Bagherinezhad, V. Karimipour, Phys.
Rev. {\bf A}, 67, 044302, (2003)], one of the receivers can cheat by
splitting the entanglement of the carrier and intercepting the
secret, without being detected. In this reply we show that a simple
modification of the protocol prevents the receivers from this kind
of cheating.
\end{abstract}
\end{titlepage}

To set up the context and the notations, it is appropriate to first
review briefly the protocol itself \cite{bk} and the basic feature
of the attack or cheating suggested in \cite{du}.
\section{The basic steps of the protocol and the cheating}

First we need the concept of a reusable secure carrier \cite{zhang},
. A Bell state like
$$|\phi^+\ra_{ab} = \frac{1}{\sqrt{2}}(|00\ra+|11\ra)_{ab},$$ shared between Alice(a) and
Bob(b) can be used as a reusable secure carrier between two parties
as follows. Alice entangles a qubit $|q\ra_1$ by the action of a
CNOT gate $C_{a1}$ (acting on the qubit $1$ and controlled by $a$),
which produces a state like
$$\frac{1}{\sqrt{2}}(|00q\ra+|11\overline{q}\ra)_{ab1}.$$ At the
destination Bob disentangles the qubit by a CNOT operation $C_{b1}$,
leaving the carrier in its original state for reusing. During the transmission the qubit has been disguised
in a highly
mixed state. \\
Any of the Bell states \be\label{bells} |\phi^{\pm}\ra_{ab} =
\frac{1}{\sqrt{2}}(|00\ra\pm|11\ra)_{ab},\h |\psi^{\pm}\ra_{ab} =
\frac{1}{\sqrt{2}}(|01\ra\pm|10\ra)_{ab}\ee
can be used as a carrier. \\

For three parties \cite{bk}, a carrier shared between Alice(a),
Bob(b) and Charlie(c) can be a GHZ state like
\begin{equation}\label{}
    |GHZ\ra:= \frac{1}{\sqrt{2}}(|000\ra+|111\ra)_{abc},
\end{equation}
or an even parity state like
\begin{equation}\label{}
    |E\ra:= \frac{1}{2}(|000\ra+|110\ra+|101\ra + |011\ra)_{abc}.
\end{equation}

Throughout \cite{bk}, the comment \cite{du} and the present reply
the subscripts $a,b$ and $c$ are used for the quibts shared by, or
the local operators acted by, Alice, Bob and Charlie respectively,
while the subscripts $1$ and $2$ are used for the qubits sent to Bob
and Charlie respectively.\\

It was shown in \cite{bk} that by suitable local operations, Alice
can send a qubit $q$ to Bob and Charlie, by entangling it to the
above carriers (hence hiding it from Eavesdroppers). In order to
share the secret between Bob and Charlie, half of the bits (the bits
in the odd rounds) were sent to Bob and Charlie, as states of the
form $|qq\ra_{12}$ which they could read without the help of each
other and the other half (the bits in the even rounds) were sent to
them in the form $\frac{1}{\sqrt{2}}(|q\ 0\ra+|\overline{q}\
1\ra_{12})$ which they could use to decipher the value of $q$ only
by their cooperation. Note that $\overline{q}=1+q\ \  {\rm mod}\ 2$.

In order to be able to send both types of states in disguised form,
Alice needs to use two types of carriers, namely the $|GHZ\ra$
carrier for the states $|q\ q\ra $ and the $|E\ra$ carrier for the
states $\frac{1}{\sqrt{2}}(|q\ 0\ra+|\overline{q}\ 1\ra)$. The
interesting point is that the two types of carriers are transformed
to each other at the end of every round by the local action of
Hadamard gates by the three parties, due to the following easily
verified property

\begin{equation}\label{three}
    H\otimes H\otimes H |GHZ\ra = |E\ra , \h H\otimes H\otimes H|E\ra = |GHZ\ra.
\end{equation}

An important property which requires careful attention is that the
carrier alternates between the above two forms regardless of the
value of the qubit $q$ which has
been sent to Bob and Charlie by Alice. \\

\begin{figure}[t]
\centering
   \includegraphics[width=10cm,height=10cm,angle=0]{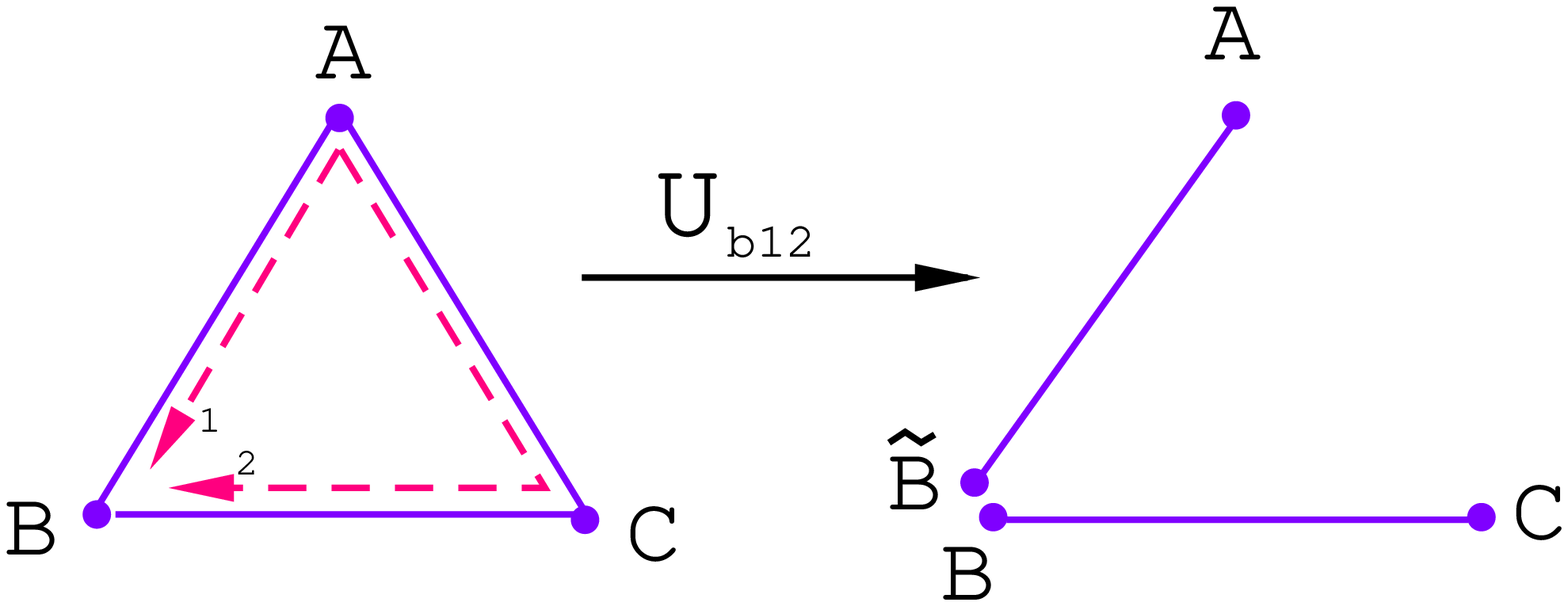}
   \caption{(Color Online) According to the comment Bob can split the three party carrier into two carriers between
   him and the other parties. The dashed arrows show the bits
   sent by Alice, the solid lines indicate the entangled states
   (carrier(s))
   shared between the parties.
   } \label{split}
\end{figure}

In \cite{du} the authors show that in the second round where a qubit
say $0$ has been encoded as $\frac{1}{\sqrt{2}}|00\ra+|11\ra_{12}$
and entangled to the carrier $|E\ra$, Bob (assuming that he has
access to the channel between Alice and Charlie) can intercept the
qubit 2 sent to Charlie (assuming that he has access to the channel
used between Alice and Charlie) and perform a suitable unitary
operation $U_{b12}$, on the state of the carrier and the two bits
$1$ and $2$, to split the carrier $|E\ra$ to two simple carriers of
the type \ref{bells}. This process is shown schematically in figure
(\ref{split}).

Let us denote by $q_2$ the qubit sent by Alice in the second round.
Bob keeps this qubit for himself and denotes it hereafter by
$\tilde{b}$, since it is now in possession of Bob and plays a role
as part of his new carriers.\\

It is important to note that the pattern of entanglement splitting
depends on the value of this qubit $q_2$ as follows (equation 3 of
the comment):

\begin{eqnarray}
  |E\ra &\lo& |\phi^+\ra_{a\tilde{b}} \otimes  |\phi^+\ra_{bc}  \h {\rm if} \ \ \ q_2=0,\\
  |E\ra &\lo& |\psi^+\ra_{a\tilde{b}} \otimes  |\psi^+\ra_{bc}  \h {\rm if} \ \ \
  q_2=1.
\end{eqnarray}

As it stands in \cite{du}, this does not harm the cheating strategy
of Bob, since as mentioned before any of the Bell states can be used
as a carrier between two parties.\

He then uses the above two pairs of entangled states for retrieving
the qubits sent by Alice on his own and sending counterfeit qubits
to Charlie in a clever way so that to avoid detection after public
announcement of subsequence of the bits.\\

What is crucial in this attack is that Bob acts by Hadamard gates on
his qubits $b$ and $\tilde{b}$ along with Alice and Charlie who are
doing the same thing at the end of each round. In this way he almost
maintains the pattern of the new carriers, which he has created in
the second round, between himself and the other two parties.\\ The
reason for "almost" is that the Hadamard operations act as follows
(equation 4 of the comment):

\begin{eqnarray}
&&  H^{\otimes 4}: |\phi^+\ra_{a\tilde{b}}\otimes |\phi^+\ra_{bc}
 \rightleftharpoons |\phi^+\ra_{a\tilde{b}}\otimes |\phi^+\ra_{bc} \\
 &&\cr
&&  H^{\otimes 4}: |\phi^-\ra_{a\tilde{b}}\otimes |\phi^-\ra_{bc}
\rightleftharpoons  |\psi^+\ra_{a\tilde{b}}\otimes
  |\psi^+\ra_{bc}.
\end{eqnarray}

Thus if the qubit $q_2$ was zero, the new two-party carriers remain
fixed at $|\phi^+\ra_{a\tilde{b}}\otimes |\phi^+\ra_{bc}$, otherwise
they alternate between the two forms $|\phi^-\ra_{a\tilde{b}}\otimes
|\phi^-\ra_{bc}$ and $|\psi^+\ra_{a\tilde{b}}\otimes
|\psi^+\ra_{bc}$. As mentioned above this does not affect his
cheating strategy, as all the Bell states are good secure carriers.

\section{Prevention of cheating}
At first sight one may argue that Alice and Charlie who are no
longer entangled after Bob's trick, can detect their new
disentangled situation (i.e. by testing a Bell inequality) and hence
detect Bob's cheating. However this test requires statistical
analysis which requires many measurements. In each measurement the
carrier collapses and will not be usable anymore.  Being in conflict
with the whole idea of reusable carrier, we do not follow this line
of argument. Instead we modify the protocol in a way which prevents
Bob's from
entanglement splitting.  \\

To this end we note that the operator $H^{\otimes 3}$ is not the
only operator which transforms the carriers $|GHZ\ra$ and $|E\ra$
into each other. Consider a unitary operator of the form
\begin{equation}\label{}
    H(\theta):=\frac{1}{\sqrt{2}}
    \left(\begin{array}{cc} e^{i\theta}& e^{-i\theta}\\ e^{i\theta}&
    -e^{-i\theta}\end{array}\right),
\end{equation}
where $\theta$ is an arbitrary parameter $\theta \in [0,2\pi)$. For
$\theta = 0$ this is the usual Hadamard operator.

Note that

\begin{equation}\label{ha}
H(\theta)|0\ra = e^{i\theta}\frac{1}{\sqrt{2}}(|0\ra+|1\ra), \h
H(\theta)|1\ra = e^{-i\theta}\frac{1}{\sqrt{2}}(|0\ra-|1\ra).
\end{equation}

A simple calculation shows that a generalization of (\ref{three}) is
possible in the following form
\begin{equation}\label{Htheta}
    H(\theta_a)\otimes H(\theta_b)\otimes H(\theta_c)|GHZ\ra =
    |E\ra\h H(\theta_a)^{-1}\otimes H(\theta_b)^{-1}\otimes
    H(\theta_c)^{-1}|E\ra=|GHZ\ra,
\end{equation}
provided that $\theta_a+\theta_b+\theta_c=0\ \ \ {\rm mod} \ \ 2\pi.
$ Therefore in the modified protocol Alice, Bob and Charlie act
alternatively by the operators $H_{\theta_a}$,\ \ $H_{\theta_b}$,
and $H_{\theta_c}$, and their inverses, on the qubits in their
possession. The angles $\theta_a, \theta_b$ and $\theta_c$ can be
announced publicly at the beginning of the protocol. We now show
that after entanglement splitting, Bob can not retain his pattern of
carriers by any operator $U_{\tilde{b}\ b}$ which he acts on his
qubits $\tilde{b}$ and $b$. We need the following\\

\textbf{Proposition}:\\

\textbf{a:} The only operator $U_{\tilde{b},b}$ which in conjunction
with $(H({\theta_a})\otimes H(\theta_c))_{ac}$ leaves invariant the
state $|\phi^+\ra_{a\tilde{b}}\otimes |\phi^+\ra_{bc}$ is the
operator $U_{\tilde{b},b} =
(H({-\theta_a})\otimes H(-\theta_c))_{\tilde{b} b}$. \\

\textbf{b:} The only operator $U_{\tilde{b},b}$ which in conjunction
with $(H({\theta_a})\otimes H(\theta_c))_{ac}$ transforms the state
$|\phi^-\ra_{a\tilde{b}}\otimes |\phi^-\ra_{bc}$ into
$|\psi^-\ra_{a\tilde{b}}\otimes |\psi^-\ra_{bc}$
 is the
operator $V_{\tilde{b},b} =
(H({\theta_a})^T\otimes H(\theta_c)^T)_{\tilde{b} b}$, where $T$ means transpose. \\

\textbf{Proof:} The proof is simply straightforward calculations. We
highlight the basic steps. Consider part \textbf{a}. We want an
operator $U_{\tilde{b}\ b}$ such that
\begin{equation}\label{}
    (H_a\otimes U_{\tilde{b}\ b} \otimes H_c)|\phi^+\ra_{a\tilde{b}}\otimes
    |\phi^+\ra_{bc}=
    |\phi^+\ra_{a\tilde{b}}\otimes |\phi^+\ra_{bc},
\end{equation}
where we use $H_a$ as an abbreviations of $H(\theta_a)_a$ and so
forth. Acting on both sides by $H_a^{-1}\otimes I\otimes I\otimes
H_c^{-1}$ we obtain
\begin{equation}\label{}
    (I_a\otimes U_{\tilde{b}\ b} \otimes I_c)|\phi^+\ra_{a\tilde{b}}\otimes
    |\phi^+\ra_{bc}= (H_a^{-1}\otimes I_{\tilde{b}}\otimes I_b\otimes H_c^{-1})
    |\phi^+\ra_{a\tilde{b}}\otimes |\phi^+\ra_{bc}.
\end{equation}

We now rearrange both sides to the convenient form
\begin{eqnarray}\label{uu}
   && (I_a\otimes I_c\otimes U_{\tilde{b}\ b})
   \left(|00\ra\otimes |00\ra +|10\ra\otimes |10\ra+|01\ra\otimes |01\ra+|11\ra\otimes
    |11\ra\right)_{ac,\tilde{b}b},\cr
    &=& (H_a^{-1}\otimes H_c^{-1}\otimes I_{\tilde{b}}\otimes I_b) \left(|00\ra\otimes |00\ra +|10\ra\otimes |10\ra+
    |01\ra\otimes |01\ra+|11\ra\otimes
    |11\ra\right)_{ac,\tilde{b}b},
\end{eqnarray}
and effect the operators $H_a^{-1}$ and $H_c^{-1}$ on the right hand
side by using (\ref{ha}). After comparing both sides in the basis
$\{|00\ra, |01\ra, |10\ra, |11\ra\}_{ac}$ we arrived at the stated
assertion, namely that $U_{\tilde{b},b} =
(H({-\theta_a})\otimes H(-\theta_c))_{\tilde{b} b}$.\\

Similar reasoning proves part $b$.\\

We now come to our main conclusion. Bob, being among the original
legitimate parties knows the values of the angles, $\theta_{a,b,c}$.
However in order to scape detection he has to apply either the
operator $U_{\tilde{b}\ b}=(H(-\theta_a)\otimes
H(-\theta_c))_{\tilde{b} b}$ or
$V_{\tilde{b},b}=(H(\theta_a)^T\otimes H(\theta_c)^T)_{\tilde{b} b}$
at the end of each round. However his choice depends on the value of
the second bit which he does not know. Without this knowledge he can
not retain the pattern of fraud carriers which he has constructed
between him and the other two parties. This then introduces errors
in half of the bits sent by Alice and received by him and Charlie,
which in subsequent public announcement of substrings of bits
reveals his cheating.  Incidentally we note that the equality
$H(-\theta)=H(\theta)^T$ holds only for $\theta=0$, that is for the
ordinary Hadamard gate.

\end{document}